\documentclass[a4paper,10pt]{article}
\usepackage{bbm}
\usepackage{opex3}
\usepackage{amsmath}
\usepackage{amssymb}
\usepackage{enumerate}
\usepackage{graphicx}
\begin{document}

\title{The geodesic form of light-ray trace in the inhomogeneous media}
\author{Kai Niu$^{1}$, Ci Song$^{1}$, Mo-Lin Ge$^{1, 2}$}
\address{$^{1}$Theoretical Physics Division, Chern Institute of Mathematics, Nankai University, Tianjin 300071, P.R.China}
\address{$^{2}$Department of Physics, College of Science, Beijing Institute of Techonology, Beijing, 100081, P.R. China}

\email{niukai@mail.nankai.edu.cn} \email{cisong@mail.nankai.edu.cn}
\email{geml@nankai.edu.cn}
\begin{abstract}
The canonical equations of the optical cloaking proposed by Shurig,
Pendry and Smith has been proved to be equivalent to the geodesic in
a 3-dimensional curved space. Carrying out the argument we extend to
the 4-dimensional Riemannian space where the extra time item appears
as the potential term in the canonical equations. The physical
meaning of the results is interpreted.
\end{abstract}
\ocis{(080.2710) Geometrical optics, inhomogeneous media; (080.2720) Geometrical optics, mathematical methods (general); (260.1180) Anisotropic media, crystal optics.}


\section{Introduction}

Recently, the theory of electromagnetic cloaking has been developed
based on coordinate transformation to achieve invisible cloaking,
which makes objects in the cloak cannot be seen from outside
\cite{first_cloaking}. This effect has been proved by ray tracing
method assuming geometrical optics limit \cite{Schurig_calculation}.
Meanwhile, a conformal mapping approach has also been used to design
a medium that can make invisible cloaking work in short-wavelength
geometrical limit \cite{leonhardt2006ocm}. After that, more specific
works have been done such as full-wave simulations of the cloaking
structures \cite{paper3}, studding on the dynamical process of
dispersive cloaking \cite{paper4} and analytical solutions on the
sensitivity of small perturbation to the cloaking
\cite{paper5,paper6}. More excitedly, the experiment of cloaking
effect has come true at microwave frequencies by the method of
metamaterial\cite{paper7}.

One of the original ideas about cloaking comes from the analogy
between a static gravitational field and a medium whose permittivity
and permeability are tensors \cite{Landau,Plebanski}. For gravity in
general relativity, 4-dimensional Riemannian geometry should work.
Meanwhile, based on the equivalent principle, the trace of the light
is also a geodesic line in the curved 4-dimensional space-time. When
the mixed time-space item $g_{0i}$ of the space-time metric equals
to zero and the time item $g_{00}$ equals to $1$ in the synchronous
reference system, we can treat 3-dimensional spatial part and time
separately. In this case, light propagates along a 3-dimensional
geodesic line. Meanwhile in the ray tracing treatment, Schurig,
Pendry and Smith \cite{Schurig_calculation} used the canonical
equations of a Hamiltonian system  to describe the light propagation
in an inhomogeneous media\cite{Kravtsov}. If the analogy between
curved space and inhomogeneous medium is valid, the propagation of
light can also be described by the geodesic equation. As all the
processes mentioned above, many theoretical works are restricted to
transformations under which time is invariant. Thus, the situations
where a cloak or an observer has an acceleration cannot be involved.
Consequently, we need to use the geodesic description.

In this letter, we develop an equivalent treatment that describes
the light-ray traces with the geodesic equation, which is solved
numerically and used to draw exactly the same pictures as drawn in
Ref. \cite{Schurig_calculation}. Then, with a proper conformal
transformation, the light-ray in the media will show a similar
action as in the static gravitational field. However, it can make an
imperfect cloaking effect which shows the light-ray extending
outside and is not a really invisible cloak. Furthermore, we
generalize the geodesic equation in 4-dimensional space-time, which
the spatial part and time part are transformed separately. In this
case, a simple example is considered as the time item $g_{00}$ of
the metric changes along the distance of the moving direction. By
choosing an appropriate parameter, the light-ray trace recurves
apparently and makes the cloaking effect imperfect similarly, which
is interpreted by non-inertial relative motion.

\section{Relation between canonical and geodesic forms of light-ray}\label{sec:analogy}

Before going to our discussion, it is useful to distinguish two
kinds of explanations of light propagation, the material form and
the geometric form \cite{Schurig_calculation}. If we interpret the
light propagation in a medium, the permittivity $\varepsilon^{ij}$
and permeability $\mu^{ij}$ of the material determine the way how
light propagates in the flat space. On the other hand, it is also
valid to interpret that light propagates along a geodesic line in a
curved vacuum space with the spacial metric $\gamma^{ij}$
\cite{Schurig_calculation}
\begin{equation}\label{eq:initial_form}
\varepsilon^{ij}=\mu^{ij}=n^{ij}=\sqrt{\gamma}\gamma^{ij},
\end{equation}
where the permittivity $\varepsilon^{ij}$ and permeability
$\mu^{ij}$ are equivalent, and $\gamma=\det(\gamma_{ij})$. For
notation here, Latin indices $i,j$ run from 1 to 3 for spatial part
and Greek indices $\mu,\nu$ run from 0 to 3 for both spatial and
time parts. In the following we denote the spatial coordinate by
$x^i$ and the space-time coordinate by $x^{\mu}$, where $x^0=ct$
describes time measured in spatial units. The sign of the curved
space-time metric takes the form $(+,-,-,-)$. The spacial metric
$\gamma_{ij}$ has the form \cite{Landau}
\begin{equation}
\gamma_{ij}=-g_{ij}+\frac{g_{0i}g_{0j}}{g_{00}},
\end{equation}
and its inverse is
\begin{equation}\label{eq:gamma_inverse}
\gamma^{ij}=-g^{ij}.
\end{equation}
Because the reference system considered by Schurig et
al\cite{first_cloaking,Schurig_calculation} is the synchronous
reference system which satisfies $g_{00}=1$ and $g_{0i}=0$. Thus, in
the condition (\ref{eq:initial_form}), we can write
\[\gamma=\det(\gamma_{ij})=-\det(g_{\mu\nu})/g_{00}=-g.\]

\subsection{Material interpretation with canonical form}

With the assumption (\ref{eq:initial_form}), the eikonal equation
\cite{Schurig_calculation} of light-ray in the inhomogeneous media
in the inertial reference can be written as
\begin{equation}\label{eq:eikonal_equation_inertial}
n^{ij}k_ik_j-\det(n^{ij})=0.
\end{equation}
Consequently, Schurig {\it et al} \cite{Schurig_calculation}
describe the light-ray in the material interpretation with the
Hamiltonian:
\begin{equation}\label{eq:original_hamiltonian}
H=f(\mathbf{x})\left[n^{ij}k_ik_j-\det{(n^{ij})}\right],
\end{equation}
where $f(\mathbf{x})$ is an arbitrary function of position
coordinates $\mathbf{x}$, and $k_i$ is wave vector. Then the
equations of motion can be given as following canonical equations:
\begin{subequations}\label{eq:canonical equations}
\begin{eqnarray}
\frac{dx^{i}}{d\tau} & = & \frac{\partial H}{\partial k_{i}},\label{eq:canonical equation dxdt}\\
\frac{dk_{i}}{d\tau} & = & -\frac{\partial H}{\partial x^{i}},\label{eq:canonical equation dkdt}
\end{eqnarray}
\end{subequations}
where $\tau$ is an arbitrary parameter describing the path. Thus a
light propagation in the cloaking material can be determined by this
procedure.

\subsection{Geometric interpretation with geodesic form}

Likewise we can also use the geometric interpretation to portray
light propagation. Actually by doing this it has following merits:
(i) Light propagating along geodesic in a curved space gives us a
vivid picture of the traces, which is easily handled by using
preliminary Riemannian geometry; (ii) With this interpretation, a
variety of techniques accumulated in the field of general relativity
may provide us potential candidates for designing cloaking
materials; (iii) Establishing this interpretation is fundamentally
important for the integrity of a theory. Based on the equivalent
principle a light ray propagates along the geodesic line and thereby
a geodesic equation should be derived from Eqs. (\ref{eq:canonical
equations}). We will show this in Section \ref{subsec:2.3}.

Firstly, the eikonal equation in the 4-dimensional space-time is
\begin{equation}\label{eq:eikonal_equation_curve_space}
k^{\mu}k_{\mu}=0.
\end{equation}
So the Hamiltonian of the light-ray can be written as
\begin{eqnarray}\label{eq:hamiltonian_4D}
H=f'(\mathbf{x})g^{\mu\nu}k_{\mu}k_{\nu},
\end{eqnarray}
where $f'(\mathbf{x})$ is another arbitrary function of position
coordinates $\mathbf{x}$. Considering the synchronous reference
system, which $g_{0i}=0$ and $g_{00}=1$, the eikonal equation is
\begin{equation}\label{eq:eikonal_equation_synchronous}
\gamma^{ij}k_ik_j-\left(\frac{\omega}{c}\right)^2=0,
\end{equation}
where $k_0=\omega/c$, $\omega$ is the angular frequency of the light
and $c$ is the speed of light. Substituting it into the geometric
Hamiltonian (\ref{eq:hamiltonian_4D}) with the unit $\omega=c=1$, we
get
\[H=f'(\mathbf{x})\left[\gamma^{ij}k_ik_j-\left(\frac{\omega}{c}\right)^2\right].\]
Comparing it with the material Hamiltonian
(\ref{eq:original_hamiltonian}), we have
\[f'(\mathbf{x})=f(\mathbf{x})\frac{\omega}{c}\sqrt{\gamma},\]
and the refractive index tensor has the form
\begin{eqnarray}\label{eq:refractive_tensor}
\varepsilon^{ij}=\mu^{ij}=n^{ij}(\mathbf{x})=\frac{\omega}{c}\sqrt{\gamma(\mathbf{x})}\gamma^{ij}(\mathbf{x}).
\end{eqnarray}
So the Hamiltonian of the light ray in the synchronous reference has
the form
\begin{equation}\label{eq:hamiltonian_synchronous_reference}
H=f(\mathbf{x})\frac{\omega}{c}\sqrt{\gamma}\left[\gamma^{ij}k_ik_j-\left(\frac{\omega}{c}\right)^2\right],
\end{equation}
which satisfies the material Hamiltonian
(\ref{eq:original_hamiltonian}) of Schurig {\it et al}
\cite{Schurig_calculation}. Our goal is, from the Hamiltonian
(\ref{eq:hamiltonian_synchronous_reference}) and the canonical
equations (\ref{eq:canonical equations}), to derive the geodesic
equation of light
\begin{equation}\label{eq:geodesic_equation}
\frac{dk^i}{d\lambda}+\Gamma^i_{jl}k^jk^l=0,
\end{equation}
where $k^i=dx^i/d\lambda$; $\lambda$ is a parameter varying along
the light traces which is also determined by Eq.
(\ref{eq:geodesic_equation}), and
$\Gamma^i_{jl}=\frac{1}{2}\gamma^{im}(\gamma_{mj,l}+\gamma_{ml,j}-\gamma_{jl,m})$
is the Christoffel symbol in 3D curved space.

\subsection{The proof process}\label{subsec:2.3}

\begin{enumerate}[a)]
\item The  arbitrary parameter $\tau$ (in Eq.
(\ref{eq:canonical equations})) and the determining parameter
$\lambda$ (in Eq.(\ref{eq:geodesic_equation})) are not same. We
assume that $\lambda$ and $\tau$ are connected by a function
$f_A(\mathbf{x})$ with arbitrary character
\begin{equation}\label{the connection between lambda and tau}
d\lambda=f_A(\mathbf{x})d\tau.
\end{equation}
\item With using the canonical equation (\ref{eq:canonical equation
dxdt}), we get (calculating details in Appendix
(\ref{appendix_A_definition})):
\begin{eqnarray}\label{eq:k^l space requairment}
k^i=\frac{dx^i}{d\lambda}=\frac{dx^i}{d\tau}\cdot\frac{d\tau}{d\lambda}=\frac{2\omega}{c}\frac{f(\mathbf{x})}{f_A(\mathbf{x})}\sqrt{\gamma}\gamma^{li}k_i.
\end{eqnarray}
If we let
$f_A(\mathbf{x})=\frac{2\omega}{c}f(\mathbf{x})\sqrt{\gamma}$.
 Eq. (\ref{eq:k^l space requairment}) satisfies $k^l=\gamma^{il}k_i$, and Hamiltonian
(\ref{eq:hamiltonian_synchronous_reference}) changes into
\begin{equation}\label{eq:space_hamiltonian_f_A}
H=\frac{f_A(\mathbf{x})}{2}\left[\gamma^{ij}k_ik_j-\left(\frac{\omega}{c}\right)^2\right].
\end{equation}
\item With calculating $dk^i/d\lambda$, we change it into the form
\[\frac{dk^i}{d\lambda}=\frac{d(\gamma^{ij}k_j)}{d\tau}\cdot\frac{d\tau}{d\lambda}=\frac{1}{f_A(\mathbf{x})}\left(\gamma^{ij}\frac{dk_j}{d\tau}+\frac{\partial \gamma^{ij}}{\partial x^m}\cdot\frac{dx^m}{d\tau}\cdot k_j\right).\]
Considering the canonical equations (\ref{eq:canonical equations}),
eikonal equation (\ref{eq:eikonal_equation_synchronous}) and the
metric property
$\gamma_{il}\gamma^{lk}_{,m}+\gamma^{lk}\gamma_{il,m}=0$, we can
finally get the geodesic equation (\ref{eq:geodesic_equation}). (See
more calculating details in Appendix
(\ref{appendix_A_demonstration}).)
\end{enumerate}

Therefore, the equivalence between the canonical form and the
geodesic form shows the following statements in the viewpoint of
mathematics. The solution of optic cloaking is the inverse problem
which is to find the metric $\gamma_{ij}$ on the given boundary
conditions as light-ray traces.

\subsection{Example of spherical cloaking}

Now we can take spherical cloaking as an example to check our
calculation. The metric inside the cloak derived from the coordinate
transformation \cite{Schurig_calculation}
\begin{equation}\label{eq:transformation}
r'=\frac{b-a}{b}r+a.
\end{equation}
Correspondingly, the metric is
\begin{equation}\label{eq:metric_sphere}
ds^2=\left(\frac{b}{b-a}\right)^2dr'^2+\left(\frac{b}{b-a}\right)^2(r'-a)^2d\theta^2+\left(\frac{b}{b-a}\right)^2(r'-a)^2\sin^2\theta
d\phi^2.
\end{equation}
So we have the material properties
\begin{eqnarray}\label{eq:material_properties_1}
\varepsilon^{r'r'}=\frac{b}{b-a}(r'-a)^2\sin\theta,\quad\varepsilon^{\theta'\theta'}=\frac{b}{b-a}\sin\theta,\quad\varepsilon^{\phi'\phi'}=\frac{b}{(b-a)}\cdot\frac{1}{\sin\theta}.
\end{eqnarray}
If the original material items
\[\varepsilon^{r'r'}_0=r'^2\sin\theta,\quad\varepsilon^{\theta'\theta'}_0=\sin\theta,\quad\varepsilon^{\phi'\phi'}_0=\frac{1}{\sin\theta}\]
are eliminated, we get
\begin{eqnarray}\label{eq:material_properties_2}
\frac{\varepsilon^{r'r'}}{\varepsilon^{r'r'}_0}=\frac{b}{b-a}\left(\frac{r'-a}{r'}\right)^2,\quad\frac{\varepsilon^{\theta'\theta'}}{\varepsilon^{\theta'\theta'}_0}=\frac{b}{b-a},\quad\frac{\varepsilon^{\phi'\phi'}}{\varepsilon^{\phi'\phi'}_0}=\frac{b}{b-a},
\end{eqnarray}
which are the same as the results in Ref.\cite{first_cloaking}.

For the calculation, the boundary conditions are given in Ref.
\cite{Schurig_calculation}
\begin{subequations}\label{eq:boundary_condition}
\begin{eqnarray}
(\mathbf{k}_1-\mathbf{k}_2)\times \mathbf{n}&=&0,\\
H(\mathbf{k}_2)&=&0,
\end{eqnarray}
\end{subequations}
\begin{figure}[h]
  \begin{minipage}[c]{0.45\columnwidth}%
                      \begin{center}
                        \includegraphics[width=0.9\columnwidth,clip]{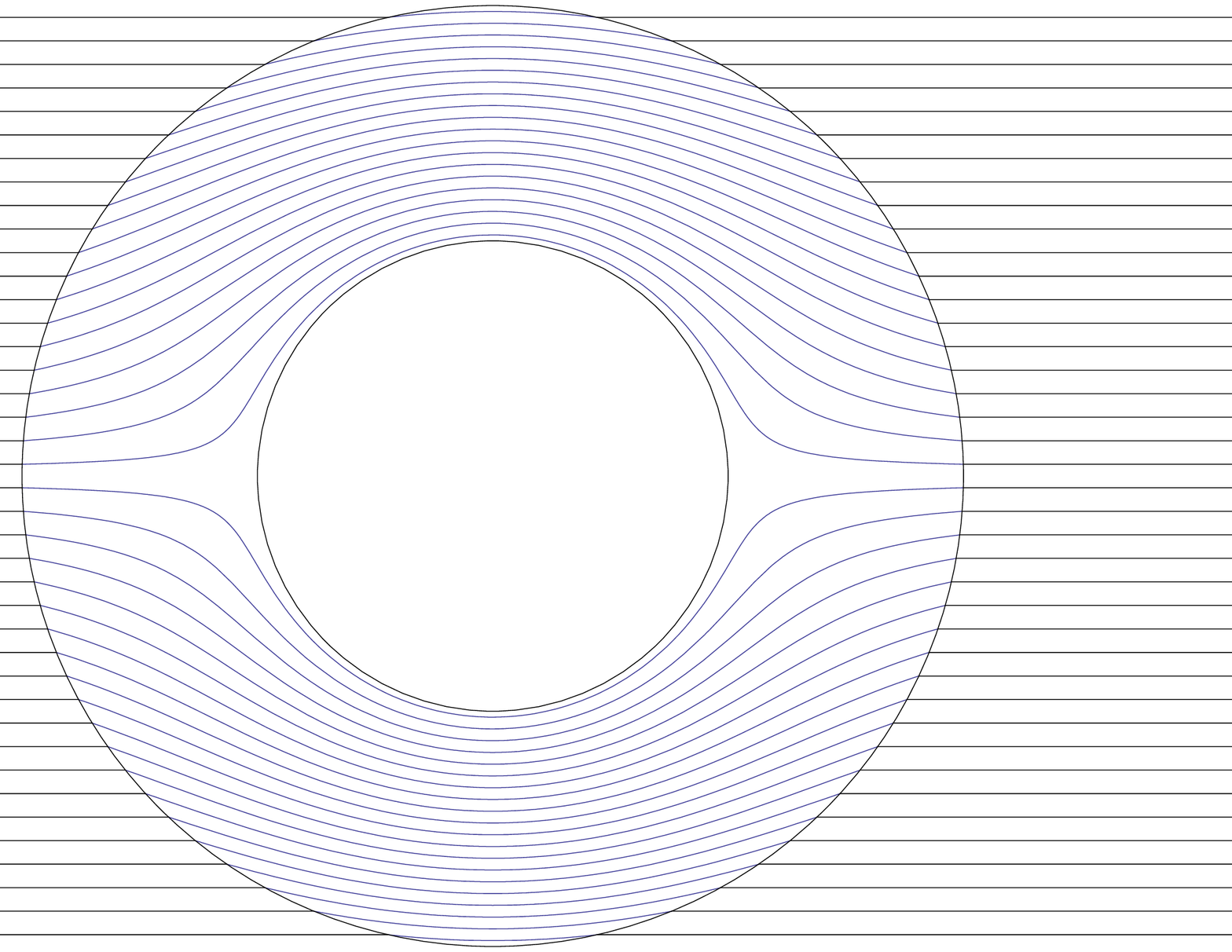}
                        \par\end{center}                         \end{minipage}%
                     \begin{minipage}[t]{0.09\columnwidth}%
                      \end{minipage}%
\begin{minipage}[c]{0.45\columnwidth}%
                        \begin{center}
                         \includegraphics[width=0.9\columnwidth]{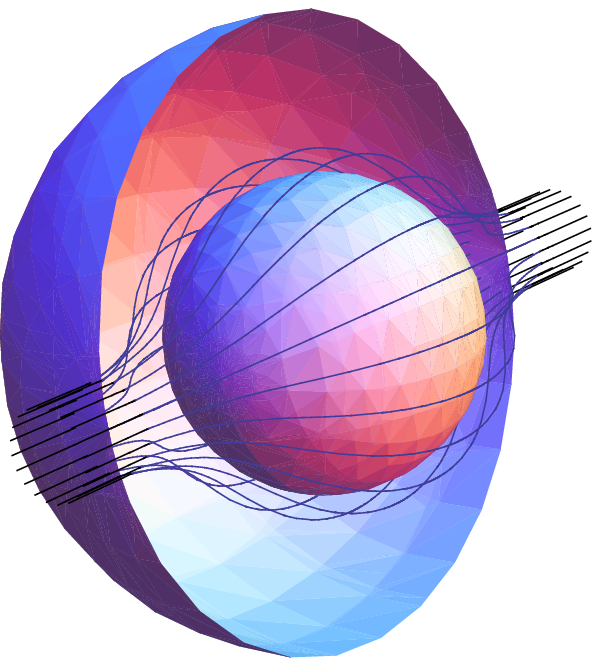}
                       \par\end{center}
\end{minipage}
  \caption{The direct calculation result of the geodesic by
   Eq. (\ref{eq:geodesic_equation}), which gives the same spherical cloaking as in Ref.  \cite{Schurig_calculation}. }
\label{fig:2d}
\end{figure}
where $\mathbf{k}_1$ is the wave vector outside of the cloak
boundary; $\mathbf{k}_2$ is inside; and $\mathbf{n}$ is the unit
normal to the boundary. By choosing $\omega/c=1$, we compute the
geodesic equation (\ref{eq:geodesic_equation}) with NDSlove of
Mathematica and draw pictures. The Fig. (\ref{fig:2d}) shows the
result. The two concentric circles and homocentric spheres are
boundaries of the cloak, and lines drawn in the pictures are
light-ray traces. We can see clearly that light-ray traces are bent
around the inside cloaking boundary, which they can still propagate
regularly out of the outside boundary. These pictures are the same
as those in Ref. \cite{Schurig_calculation}.

\section{The effect of conformal transformation on 3-dimensional metric}\label{sec:conformal}
\subsection{A trial on 4-dimensional extending}
In the previous section, we establish the geodesic description of
the light propagation in an invisible cloaking. In this section, we
want to extend the analogy between the inhomogeneous media
$\varepsilon^{ij}=\mu^{ij}=n^{ij}$ and the curved space from three
dimensions spacial space to four dimensions space-time. As in the
Ref. \cite{Landau} Landau and Lifshitz proved a static gravitational
field plays the role of a medium with permittivity and permeability
which satisfy
$\varepsilon^{ij}=\varepsilon_0\gamma^{ij}/\sqrt{g_{00}},\quad\mu^{ij}=\mu_0\gamma^{ij}\sqrt{g_{00}}$
as follows.

For Maxwell's equations in the curved 4-dimensional space-time
\begin{subequations}\label{eq:Maxwell_eq_curve_components}
\begin{eqnarray}
\label{eq:B_Gauss_Law_curve_components}\frac{1}{\sqrt{\gamma}}\partial_i(\sqrt{\gamma}B^i)&=&0,\\
\label{eq:Faraday_Law_curve_components}\frac{\partial B^i}{\partial
t}+\frac{1}{\sqrt{\gamma}}\epsilon^{ijk}\partial_jE_k&=&0,\\
\label{eq:D_Gauss_Law_curve_components}\frac{1}{\sqrt{\gamma}}\frac{\partial}{\partial
x^i}(\sqrt{\gamma}D^i)&=&\rho,\\
\label{eq:Ampere_Law_curve_components}-\frac{\partial D^i}{\partial
t}+\frac{1}{\sqrt{\gamma}}\epsilon^{ijk}\partial_jH_k&=&J^i;
\end{eqnarray}
\end{subequations}
and their apparently covariant form
\begin{subequations}\label{eq:Maxwell_eq_curve_covar}
\begin{eqnarray}
\label{eq:Maxwell_curve_1st}\partial_{\lambda}F_{\mu\nu}+\partial_{\nu}F_{\lambda\mu}+\partial_{\mu}F_{\nu\lambda}&=&0,\\
\label{eq:Maxwell_curve_2st}\frac{1}{\sqrt{-g}}\partial_{\alpha}(\sqrt{-g}G^{\alpha\beta})&=&j^{\beta}.
\end{eqnarray}
\end{subequations}
Thus, comparing with two forms of Maxwell's equations
(\ref{eq:Maxwell_eq_curve_components},\ref{eq:Maxwell_eq_curve_covar}),
we get
\begin{eqnarray*}\label{eq:em_tensors_vectors}
\left\{\begin{array}{rcl}
F_{ij}&=&-c\sqrt{\gamma}\epsilon_{ijk}B^k\\
F_{0i}&=&E_i\\
G^{i0}&=&\frac{c}{\sqrt{g_{00}}}D^i\\
G^{ij}&=&-\frac{1}{\sqrt{-g}}\epsilon^{ijk}H_k
\end{array}\right.\ \ \ \ \ \ \ \ \ \
\left\{
\begin{array}{rcl}
B^i&=&-\frac{1}{2c\sqrt{\gamma}}\epsilon^{ijk}F_{jk}\\
E_i&=&F_{0i}\\
D^i&=&\frac{\sqrt{g_{00}}}{c}G^{i0}\\
H_i&=&-\frac{\sqrt{-g}}{2}\epsilon_{ijk}G^{jk}
\end{array}\right.
\end{eqnarray*}
where $\epsilon^{ijk}$ and $\epsilon_{ijk}$ are the Levi-Civita
symbol. Considering the vacuum media in the curved space-time, we
have the relation
\begin{equation}\label{eq:relation_FG}
G^{\alpha\beta}=g^{\alpha\mu}g^{\beta\nu}F_{\mu\nu}.
\end{equation}
Thus, we have
\[\left\{
\begin{array}{rcl}
D^i&=&[\varepsilon_0\sqrt{g_{00}}(g^{0i}g^{0j}-g^{00}g^{ij})]E_j+[-c\varepsilon_0\sqrt{-g}g^{ij}g^{0k}\epsilon_{jkl}]B^{l},\\\\
B^i&=&[\frac{c\mu_0}{\sqrt{-g}}g_{0j}g_{kl}\epsilon^{ijk}]D^l+[\frac{\mu_0}{\sqrt{g_{00}}}\cdot\frac{1}{2\gamma}g_{jl}g_{km}\epsilon^{ijk}\epsilon^{lmn}]H_n.
\end{array}\right.
\]
where we set $\varepsilon_0=\mu_0=1/c$. If the curved space-time
satisfies $g_{0i}=0$ and $g_{ij}=-\gamma_{ij}$, then we have
\begin{equation}\label{eq:Definition_em_No_vacuum_curve}
D^i=\varepsilon_0\frac{\gamma^{ij}}{\sqrt{g_{00}}}E_j,\quad
B^i=\mu_0\frac{\gamma^{ij}}{\sqrt{g_{00}}}H_j.
\end{equation}
Comparing it with the inhomogeneous media in the flat space-time,
\begin{equation}\label{eq:Definition_em_No}
D^i=\varepsilon^{ij}E_j,\quad B^i=\mu^{ij}H_j,
\end{equation}
we have the result
\begin{eqnarray}\label{eq:curve_media}
\varepsilon^{ij}=\varepsilon_0\frac{\gamma^{ij}}{\sqrt{g_{00}}},\quad\mu^{ij}=\mu_0\frac{\gamma^{ij}}{\sqrt{g_{00}}}.
\end{eqnarray}

\subsection{Conformal transformation}

Thus, We will show how to design a media satisfies the condition
(\ref{eq:curve_media}). Our first trial is to make a conformal
transformation
\begin{equation}\label{eq:conformal_transformation}
\gamma^{ij}\rightarrow \gamma'^{ij}=\sigma(\mathbf{x})\gamma^{ij},
\end{equation}
Substituting it into the eikonal equation
(\ref{eq:eikonal_equation_synchronous}), we get the Hamiltonian
\[H=f'(\mathbf{x})\left[\sigma(\mathbf{x})\gamma^{ij}k_ik_j-\left(\frac{\omega}{c}\right)^2\right].\]
Similarly, comparing it with the material Hamiltonian
(\ref{eq:original_hamiltonian}), we have
\begin{eqnarray*}
f'(\mathbf{x})=f(\mathbf{x})\frac{\omega}{c\sigma}\sqrt{\frac{\gamma}{\sigma}},
\end{eqnarray*}
and the refractive index tensor has the form
\begin{eqnarray}\label{eq:refractive_tensor_conformal}
\varepsilon^{ij}=\mu^{ij}=n^{ij}=\frac{\omega}{c}\sqrt{\gamma'}\gamma'^{ij}=\frac{\omega}{c}\sqrt{\frac{\gamma}{\sigma(\mathbf{x})}}\gamma^{ij}.
\end{eqnarray}
Thus the Hamiltonian of the light ray in the synchronous reference
with a conformal transformation (\ref{eq:conformal_transformation})
has the form
\begin{eqnarray*}\label{eq:hamiltonian_synchronous_reference_conformal}
H=f(\mathbf{x})\frac{\omega}{c\sigma}\sqrt{\frac{\gamma}{\sigma}}\left[\sigma(\mathbf{x})\gamma^{ij}k_ik_j-\left(\frac{\omega}{c}\right)^2\right].
\end{eqnarray*}
If we want the definition of $k^i=dx^i/d\lambda$ satisfying
$k^i=\gamma'^{ij}k_j=\sigma(\mathbf{x})\gamma^{ij}k_j$, it should be
\begin{eqnarray*}
f_A(\mathbf{x})=\frac{2\omega}{c\sigma(\mathbf{x})}\sqrt{\frac{\gamma}{\sigma(\mathbf{x})}}f(\mathbf{x}).
\end{eqnarray*}
So the Hamiltonian (\ref{eq:original_hamiltonian}) changes to
\begin{equation}\label{eq:conformal_space_hamiltonian}
H=\frac{f_A(\mathbf{x})}{2}\left[\sigma(\mathbf{x})\gamma^{ij}k_ik_j-\left(\frac{\omega}{c}\right)^2\right].
\end{equation}
Calculating $dk^i/d\lambda$ with the canonical Eqs.
(\ref{eq:canonical equations}) and the Hamiltonian
(\ref{eq:conformal_space_hamiltonian}), we get the new geodesic
equation in 3D space
\begin{equation}\label{eq:geodesic_equation_conformal}
\frac{dk^i}{d\lambda}+\Gamma'^i_{st}k^sk^t=0,\ \ \ \ \ \ \
\Gamma'^i_{st}=\Gamma^i_{st}+\left(\frac{\partial
\ln\sqrt{|\sigma(\mathbf{x})|}}{\partial
x^j}\gamma^{ij}\gamma_{ts}-\delta^i_t\frac{\partial
\ln|\sigma(\mathbf{x})|}{\partial x^s}\right).
\end{equation}
The details are in Appendix (\ref{appendix_B}).

\subsection{Physical interpretation}

Comparing the similarity between the curved space-time
(\ref{eq:curve_media}) with the refractive index $n^{ij}$ and
conformal transformation (\ref{eq:refractive_tensor_conformal}), we
get the analog
\begin{eqnarray}
\sigma(\mathbf{x})=-g(\mathbf{x})=g_{00}\cdot\gamma.
\end{eqnarray}
So the media with the condition
(\ref{eq:refractive_tensor_conformal}) can describe an artificial
static gravitational field for the light-ray. Meanwhile, we can also
make the analog
\begin{equation}\label{eq:sigma_g_00}
\sigma(\mathbf{x})=g_{00}
\end{equation}
for the next section with different physical meaning. It is similar
to the 4-D coordinate transformation but totally different, which
will be discussed later.

Considering the Hamiltonian (\ref{eq:conformal_space_hamiltonian}),
if the function with arbitrary property $f_A(\mathbf{x})$ satisfies
\begin{equation}\label{eq:physical_meaning_f_A}
f_A(\mathbf{x})=\frac{1}{m\sigma(\mathbf{x})}\ \ \ \ \ \ \ \
(m=\frac{\hbar\omega}{c^2}),
\end{equation}
the Hamiltonian (\ref{eq:conformal_space_hamiltonian}) recasts to
\begin{eqnarray}\label{eq:physical_meaning_hamiltonian}
H=K(\mathbf{k})+V(\mathbf{x})=\frac{\gamma^{ij}k_ik_j}{2m}-\frac{1}{2\sigma(\mathbf{x})}\left(\frac{c}{\hbar}\right)^2m.
\end{eqnarray}
Here, the photon in the media is massless, but the speed is less
than $c$. So we can treat it as a ``non-relative particle'' with an
equivalent mass $m$. The Hamiltonian
(\ref{eq:physical_meaning_hamiltonian}) has a kinetic energy
$K(\mathbf{k})$ with a inertia mass $m$ and a potential
$V(\mathbf{x})$ with a ``gravitational'' mass $m$, where the inertia
mass and the ``gravitational'' mass equal to each other. Taking
account of the  principle of equivalence, it can be viewed as a
particle $m$ accelerating in a curved 3D space with spacial metric
$\gamma_{ij}$. The ``Newtonian equation '' is
\begin{equation}\label{eq:acceleration}
\dot{\mathbf{v}}=\frac{1}{2\sigma(\mathbf{x})^2}\left(\frac{c}{\hbar}\right)^2\frac{\partial
\sigma(\mathbf{x})}{\partial
\mathbf{x}}-\frac{k_ik_j}{2m^2}\left(\frac{\partial
\gamma^{ij}(\mathbf{x})}{\partial \mathbf{x}}\right),
\end{equation}
where the first term is generated by the potential $V(\mathbf{x})$
and the second is contributed by kinetic energy in the  curved
space.

Consequently, with the given Hamiltonian
(\ref{eq:physical_meaning_hamiltonian}), we can write the Poisson
brackets
\[\{F,G\}=\frac{\partial F}{\partial k_n}\cdot\frac{\partial
G}{\partial x^n}-\frac{\partial F}{\partial x^n}\cdot\frac{\partial
G}{\partial k_n},\]
and the following results
\begin{subequations}
\begin{eqnarray}
\{x^i,k_j\}&=&-\delta^i_j,\\
\{x^i,k^j\}&=&-\gamma^{ij}\\
\{H,x^i\}&=&\frac{\partial H}{\partial k_n}=\frac{\gamma^{ij}k_j}{m},\\
\{H,k_i\}&=&-\frac{\partial H}{\partial
x^i}=-\frac{k_sk_t}{2m}\left(\frac{\partial \gamma^{st}}{\partial
x^i}\right)-\frac{m}{2\sigma(\mathbf{x})^2}\left(\frac{c}{\hbar}\right)^2\frac{\partial
\sigma(\mathbf{x})}{\partial x^i}.
\end{eqnarray}
\end{subequations}
It may be viewed as a classical model for the light propagation in
the medium with $\varepsilon^{ij}=\mu^{ij}=n^{ij}$.

Here, we need to notice that the conformal transformation with
analog (\ref{eq:sigma_g_00}) can not give a real 4-dimensional
curved coordinates. The gravitational ``red-shift'' effect does not
occur and the light-ray trace is not the geodesic line in the
4-dimensional curved space-time. The conformal transformation in
3-dimensional can't be given by the coordinate transformation. The
real 4-dimensional coordinate transformation will be given in the
next section. So the analog (\ref{eq:sigma_g_00}) cannot show a real
4-dimensional transformation as the gravitational field or any
accelerating systems.

\subsection{Example of spherical cloaking with conformal transformation}

If we still take the coordinate transformation
(\ref{eq:transformation}) and choose $\sigma(\mathbf{x})$ properly,
we can make sure that the light can not go into the cloak region
$r<a$.

\begin{figure}[h]
  \begin{minipage}[c]{0.45\columnwidth}%
                      \begin{center}
                        \includegraphics[width=0.8\columnwidth,clip]{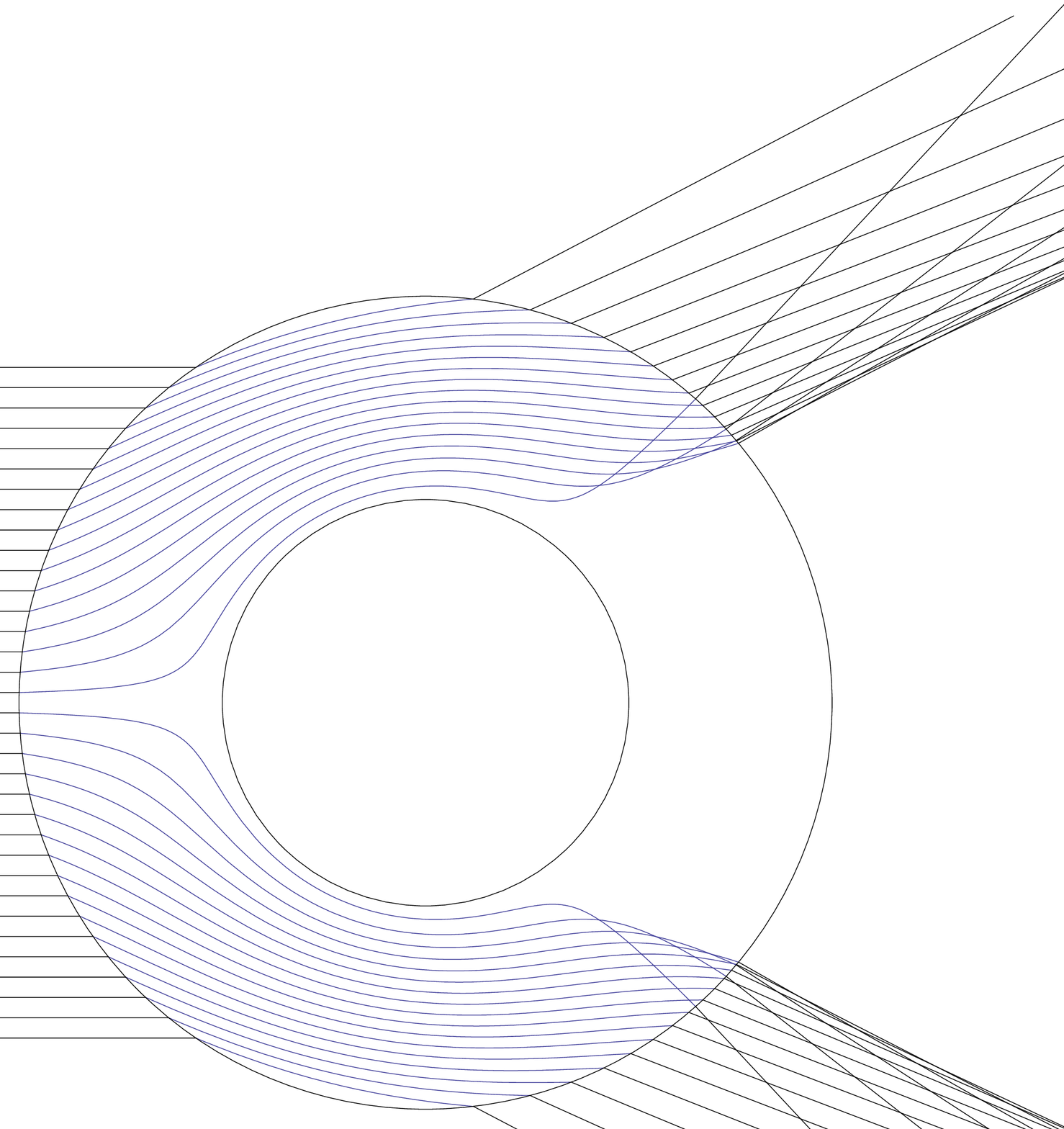}
                        \par\end{center}                        \end{minipage}%
                     \begin{minipage}[t]{0.09\columnwidth}%
                      \end{minipage}%
\begin{minipage}[c]{0.45\columnwidth}%
                        \begin{center}
                         \includegraphics[width=\columnwidth]{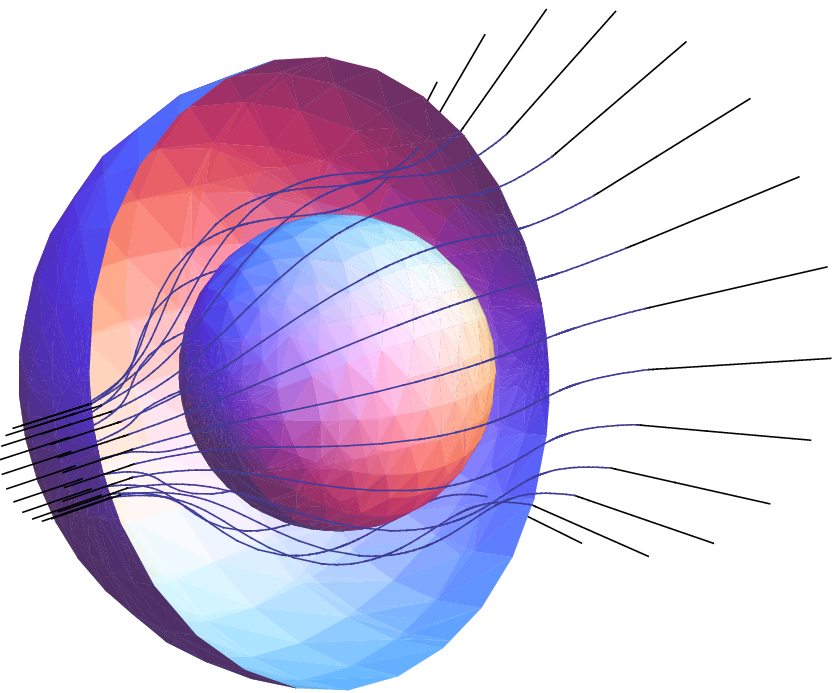}
                       \par\end{center}
\end{minipage}
  \caption{Adding $\sigma(\mathbf{x})$ cloak: the adding $\sigma(\mathbf{x})$ term bends the light-rays and can make sure the light does not go into the
cloaked region $r<a$.}
  \label{fig:2d_add_g00}
\end{figure}

By choosing $\sigma(\mathbf{x})=1+\frac{b-a}{2b(r-a)}$ and
calculating Eq.  (\ref{eq:geodesic_equation_conformal}), we  draw
Fig.  (\ref{fig:2d_add_g00}). From the picture it can be seen that
the light ray is bent towards outside. Here we choose a proper
$\sigma(\mathbf{x})$. Thus, the cloak can still be preserved to some
extend. In general relativity \cite{Landau}, sometimes $g_{00}$ has
a direct relationship with Newtonian potential. So
$\sigma(\mathbf{x})=g_{00}(\mathbf{x})=1+\frac{b-a}{2b(r-a)}$
represents a field having repulsion force, which causes the light
ray extending outside.

\section{The cloaking system with 4-dimensional coordinate transformation}\label{sec:4D}

At the beginning, we discuss the problem in the synchronous
reference, which $g_{00}=1$ and $g_{0i}=0$. Like Ref.
\cite{Schurig_calculation}, the analogy between the inhomogeneous
media and the 3D curved space has been shown and  the time $x^0=ct$
has not been transformed. Additionally, we try to give a trial on
4-dimensional cloaking system by conformal transformation. However,
it is not a real 4-dimensional cloaking system. More generally, we
will discuss how to make the time item $g_{00}=g_{00}(\mathbf{x})$ a
function with variant spacial coordinate $\mathbf{x}$, which extends
a real analogy in Ref. \cite{Schurig_calculation}. Meanwhile, it can
be interpreted as the cloaking system or an observer has an
accelerating motion.

\subsection{The light-ray's change under time transformation}
For the eikonal equation (\ref{eq:eikonal_equation_curve_space})
given previously, we have its form under the condition $g_{00}$
changes to
\[\gamma^{ij}k_ik_j-g^{00}(k_0)^2=0.\]
So we have its Hamiltonian
\begin{eqnarray}\label{eq:hamiltonian_4D_g_00}
H=f'(\mathbf{x})\left[\gamma^{ij}k_ik_j-g^{00}(k_0)^2\right],\quad
k_0=\frac{\omega_0}{c}.
\end{eqnarray}
With the same process of demonstration, we can prove the light-ray
traces are consistent  with the geodesic lines in 4-dimensional
curved space with $g_{0i}=0$ (calculating details in Appendix
(\ref{appendix_C_proof}))
\begin{eqnarray}\label{eq:geodesic_4D}
\frac{dk^i}{d\lambda}+\Gamma^i_{jl}k^jk^l+\Gamma^i_{00}(k^0)^2=\frac{dk^i}{d\lambda}+\left(\Gamma^i_{jl}+\gamma_{jl}\gamma^{im}\frac{\partial
\ln\sqrt{g_{00}}}{\partial x^m}\right)k^jk^l=0,
\end{eqnarray}
where the third item on the left side of the equation
(\ref{eq:geodesic_4D}) above recurves the geodesic line in
3-dimensional space, and whatever $\sigma(\mathbf{x})$ is, the
equations are totally different from the equations
(\ref{eq:geodesic_equation_conformal}) (see the discussion in
Appendix (\ref{appendix_C_compare})). The Christoffel symbol
$\Gamma^i_{00}$ and $\Gamma^i_{jl}$ are components of the 4-D
Christoffel symbol $\Gamma^{\mu}_{\rho\lambda}$. However, in the
condition $g_{0i}=0$, $\Gamma^i_{jl}$ are also 3-D Christoffel
symbol.

Here, we need to notice the material Hamiltonian
(\ref{eq:original_hamiltonian}) does not fit to the condition here.
Because the Hamiltonian (\ref{eq:original_hamiltonian}) describes
the light-ray in the media in a synchronous reference system. The
reference here, however, is not synchronous any more, which time
item $g_{00}$ changes along the path in the spacial space. In
consequence of this, we need to redesign the material property for
cloaking optic device.

\subsection{Cloaking design and ``red-shift''}
Now, we consider how the refractive index changes and how the
changes will affect the cloaking system under the time
transformation. As in the synchronous reference, we take time and
space transformation independently. So
\[\Lambda_{\alpha'}^\alpha=\frac{\partial
x^{\alpha}}{\partial
x^{\alpha'}}\quad\textrm{and}\quad\Lambda_{0'}^{i}=0,\quad\Lambda_0^{0'}=\psi(\mathbf{x}),\]
which shows the time part transformation only depend on the spatial
coordinate. Thus, we get
\[dt'=\psi(\mathbf{x})dt\quad\textrm{and}\quad t'=\psi(\mathbf{x})t.\]
Because of $g_{00}dt'^2=1\cdot dt^{2}$, we can write
\begin{eqnarray}\label{eq:g_00}
\psi(\mathbf{x})=\frac{1}{\sqrt{g_{00}}}.
\end{eqnarray}

Now, we take Faraday's law of induction as an example. With
$B^i(\mathbf{x},t)=\mu^{ij}(\mathbf{x})H_j(\mathbf{x})e^{i\omega
t}$, the equation has the form
\begin{align}\label{eq:mu1}
\epsilon^{ijk}\frac{\partial E_{k}(\mathbf{x})}{\partial
x^{j}}+\mu^{ij}(\mathbf{x})H_{j}\frac{\partial e^{i\omega
t}}{\partial t} & =0.
\end{align}
For the 4-dimensional space-time transformation, the equation turns
out to be
\begin{align}\label{eq:mu2}
\epsilon^{i'j'k'}\left[\frac{E'_{k'}(\mathbf{x}')}{\partial
x'^{j'}}\right] &
+\left[\frac{1}{\det(\Lambda^{i'}_i)\sqrt{g_{00}}}\Lambda^{i'}_i\Lambda^{j'}_j\mu^{ij}(\mathbf{x})\right]H'_{j'}(\mathbf{x}')\frac{\partial
e^{i\omega\sqrt{g_{00}}t'}}{\partial t'}=0,
\end{align}
Comparing Eqs. (\ref{eq:mu1}) with (\ref{eq:mu2}),  we can see that
$\mu^{ij}$ transforms to
\begin{align}\label{eq:cloaking_device}
\mu'^{i'j'}(\mathbf{x}')=\frac{1}{\det(\Lambda^{i'}_i)\sqrt{g_{00}}}\Lambda^{i'}_i\Lambda^{j'}_j\mu^{ij}(\mathbf{x})
\end{align}
and $\omega$ transforms to
\begin{align}\label{eq:red-shift}
\omega' & =\omega\sqrt{g_{00}}.
\end{align}
The Amp\'ere's circuital law without electric current has the same
transformation properties. Thus, we design the cloaking device which
has the property of
\begin{equation}\label{eq:cloaking_device_4D}
\varepsilon^{ij}=\mu^{ij}=n^{ij}=\sqrt{\frac{\gamma}{g_{00}}}\gamma^{ij}
\end{equation}
and the light for the cloaking effect will show a ``red-shift'' like
(\ref{eq:red-shift}).

\subsection{Physical Interpretation}

As a physical example, we consider the following situation. Assuming
the cloak  moves in a constant acceleration $\mathcal{A}$ relative
to an observer and due to the equivalence principle, it is equal to
say there is a constant gravity in the space-time whose metric is
the so-called Rindler metric \cite{uni_acce}
\begin{equation}
ds^{2}=(1+\frac{\mathcal{A}z}{c^2})^{2}dt^{2}-dx^{2}-dy^{2}-dz^{2}.\label{eq:metric}
\end{equation}

Based on this time transformation, we see that in this reference
$\epsilon^{ij}$ and $\mu^{ij}$ in the cloak change to be
$\epsilon^{ij}=\mu^{ij}=\sqrt{\gamma}\gamma^{ij}/\sqrt{g_{00}}$,
where $g_{00}=\left(1+\frac{\mathcal{A}z}{c^2}\right)^{2}$. The
Hamiltonian can be written as
\begin{equation}\label{4DH}
H=\frac{1}{2}\left[\gamma_{ij}k^{i}k^{j}-\left(1+\frac{\mathcal{A}z}{c^2}\right)^{2}\left(\frac{\omega'}{c}\right)^{2}\right].
\end{equation}
\begin{figure}[h]
\begin{minipage}[t]{0.45\columnwidth}%
\begin{center}
\includegraphics[width=0.9\columnwidth,clip]{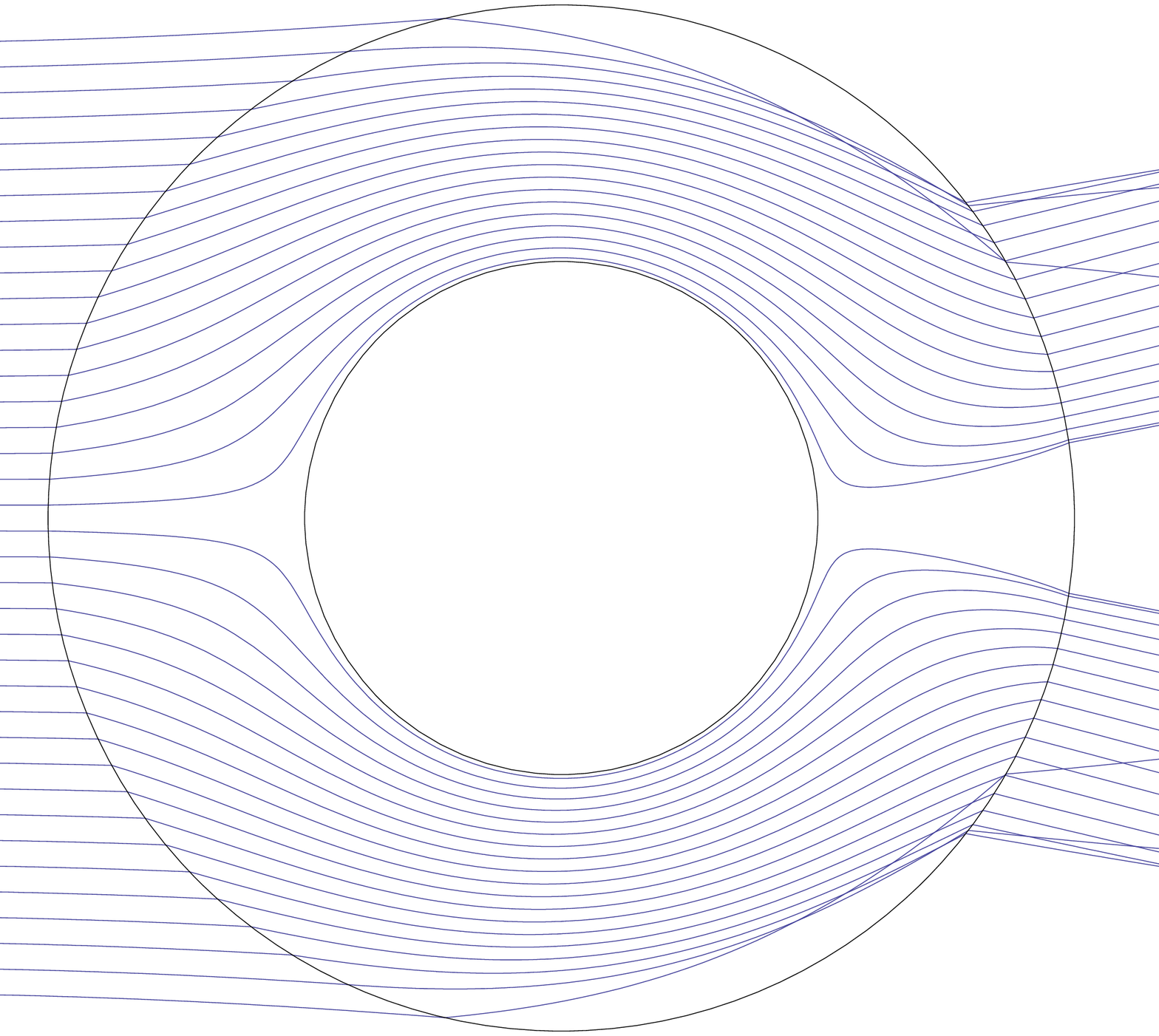}
\par\end{center}
\end{minipage}%
\begin{minipage}[t]{0.09\columnwidth}%
\end{minipage}
\begin{minipage}[t]{0.45\columnwidth}%
\begin{center}
\includegraphics[width=1\columnwidth]{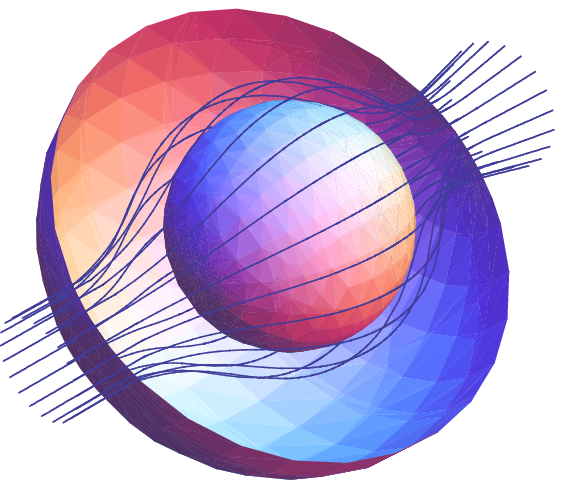}
\par\end{center}
\end{minipage}
\caption{Cloak with a time transformation: The acceleration we
choose to draw the picture is very large
$\mathcal{A}=0.11\times(3\times 10^8)^2\text{m}/\text{s}^2$, when
$\mathcal{A}$ decreases  to one tenth, the light-rays observed in
the acceleration reference will be analogous to the light-rays in
the original inertia reference.} \label{Flo:2d_add_g00-1}
\end{figure}

In the new metric (\ref{eq:metric}), we still take the
transformation (\ref{eq:transformation}). Parameters for the
calculation are $a=1$m, $b=2$m, $\mathcal{A}=0.11\times(3\times
10^8)^2\text{m}/\text{s}^2$, and the results are shown in Fig.
(\ref{Flo:2d_add_g00-1}). If $\mathcal{A}$ decreases to one tenth,
the light-rays  observed in the acceleration reference will be
analogous to the light-rays in the original inertia reference, but
meanwhile $\mathcal{A}$ is still very large. So if we are in a
reference which has an acceleration that is the same as the
acceleration of a body on the earth surface, it is impossible for us
to see the changes of the light  through the cloak compared to the
original cloak in an inertia reference.

The curvature of this space-time is not zero. Actually the scalar
Ricci curvature reads
\begin{align*}
R&=\frac{4a\mathcal{A}\cos\theta(-b+a)^{2}}{(-r+a)^{2}b^{2}\left(1+\mathcal{A}r\cos\theta\right)}.
\end{align*}
However, the curvature of space-time is zero when we just consider
3D space coordinate transformations.

\section{Conclusion}

In this paper, we establish the relationship between the canonical
equations and the geodesic equation describing the light
propagation, and use the latter form to obtain the light-ray trace
in the cloaking system. In the discussion, it is easy to show that
equations of both forms can derive the same light-ray traces in an
invisible cloaking system. Then we generalize the geodesic equation
with a 3D conformal transformation in metric. By choosing a specific
3D coordinate transformation and the conformal transforation item
$\sigma(\mathbf{x})$ properly, we can also make sure that the
light-ray do not go into cloak region $r<a$. In addition, we give
the physical interpretation of the conformal transformation that the
photon in the inhomogeneous media can be viewed as a particle with
an equivalent mass $m$ accelerating in a curved 3D space. Finally,
we make a 4D coordinate transformation which the time item $g_{00}$
is variable and the ``red-shift'' effect and accelerating cloak can
be seen.

\section*{Acknowledgment}

We are most grateful to  D. Schurig for his generosity on sharing
his computer program with us. Meanwhile, we also thank M.G. Hu for
his helpful discussion and checking the manuscript carefully. M.L.
Ge thanks to Prof. C.H. Gu for enlightening discussion in
mathematics. This work was supported in part by NSF of China (Grants
No. 10605013), and by Liu- Hui Center for Applied Mathematics
through the joint project of Nankai University and Tianjin
University.

\begin{appendix}
\section{The demonstration of equivalence between canonical equations and geodesic}\label{appendix_A}
\subsection{The equivalence of $k^l$'s definition}\label{appendix_A_definition}
Firstly, we have the definition of wave vector with upper index
\begin{equation}\label{eq:A_k^i_space definition}
k^{i}=\frac{dx^i}{d\lambda},
\end{equation}
where $\lambda$ is determined by the geodesic Eq.
(\ref{eq:geodesic_equation}) \cite{Landau}. And we also know that
the $k^l$, as a contravariant vector, satisfies
\begin{equation}\label{eq:A_k^i_vector definition}
k^i=\gamma^{il}k_l.
\end{equation}
We need these two forms to be equivalent.

So by using Eq. (\ref{eq:A_k^i_space definition}) we have
\begin{eqnarray*}
k^i=\frac{dx^i}{d\tau}\cdot\frac{d\tau}{d\lambda}=\frac{\partial
H}{\partial
k_i}\cdot\frac{1}{f_A(\mathbf{x})}=\frac{2\omega}{c}\frac{f(\mathbf{x})}{f_A(\mathbf{x})}\sqrt{\gamma}\gamma^{il}k_l
\end{eqnarray*}
If we let the arbitrary functions $f_A(\mathbf{x})$ and
$f(\mathbf{x})$ satisfy the relation
\begin{equation}
f_A(\mathbf{x})=\frac{2\omega}{c}\sqrt{\gamma}f(\mathbf{x}),
\end{equation}
two forms of $k^i$ (\ref{eq:A_k^i_vector
definition},\ref{eq:A_k^i_space definition}) are equivalent.

\subsection{The deduction from canonical equations to geodesic
equation}\label{appendix_A_demonstration}

With the canonical Eqs. (\ref{eq:canonical equations}) and the Eq.
(\ref{eq:A_k^i_vector definition}), we have

\begin{eqnarray*}
\frac{dk^{i}}{d\lambda} & =&\frac{d\left(\gamma^{ij}k_{j}\right)}{d\tau}\cdot\frac{d\tau}{d\lambda}\\
 & =&\left(\frac{d\gamma^{ij}}{d\tau}k_{j}+\gamma^{ij}\frac{dk_{j}}{d\tau}\right)\cdot\frac{1}{f_A(\mathbf{x})}\\
 & =&\left(\frac{\partial \gamma^{ij}}{\partial x^{s}}\frac{dx^{s}}{d\tau}k_{j}-\gamma^{ij}\frac{\partial H}{\partial x^{j}}\right)\cdot\frac{1}{f_A(\mathbf{x})}\\
 & =&\frac{\partial \gamma^{ij}}{\partial x^{s}}k^{s}k_{j}-\gamma^{ij}\frac{1}{2}\frac{\partial \gamma^{lm}}{\partial x^{j}}k_{l}k_{m}-\gamma^{ij}\frac{1}{2}\frac{df_{A}(x)}{dx^{j}}\left[\gamma^{lm}k_{l}k_{m}-\left(\frac{\omega}{c}\right)^2\right]\cdot\frac{1}{f_A(\mathbf{x})}\\
\text{combining with} && \gamma^{lm}k_{l}k_{m}-\left(\frac{\omega}{c}\right)^2=0\\
& =&\gamma_{jt}\frac{\partial \gamma^{ij}}{\partial x^{s}}k^{s}k^{t}-\frac{1}{2}\gamma_{ls}\gamma_{mt}\gamma^{ij}\frac{\partial \gamma^{lm}}{\partial x^{j}}k^{s}k^{t}-0\\
 & =&\left(\frac{\partial\left(g_{jt}g^{ij}\right)}{\partial x^{s}}-g^{ij}\frac{\partial g_{jt}}{\partial x^{s}}\right)k^{s}k^{t}-\frac{1}{2}g^{ij}g_{mt}\left(\frac{\partial \left(g_{ls}g^{lm}\right)}{\partial x^{j}}-g^{lm}\frac{\partial g_{ls}}{\partial x^{j}}\right)k^{s}k^{t}\\
 & =&-\gamma^{ij}\left(\frac{1}{2}\frac{\partial \gamma_{jt}}{\partial x^{s}}+\frac{1}{2}\frac{\partial \gamma_{jt}}{\partial x^{s}}-\frac{1}{2}\frac{\partial \gamma_{st}}{\partial x^{j}}\right)k^{s}k^{t}\\
 & =&-\Gamma_{st}^{i}k^{s}k^{t}.
\end{eqnarray*}
So we get the geodesic Eq. (\ref{eq:geodesic_equation}).

\section{the change of geodesic equation under 3-dimensional conformal
transformation}\label{appendix_B}

Under the conformal transformation
(\ref{eq:conformal_transformation}), we can derive another geodesic
equation (\ref{eq:geodesic_equation_conformal}) from the canonical
equations (\ref{eq:canonical equations}).
\begin{align*}
\frac{dk^{i}}{d\lambda} & =\frac{d\left(\sigma(\mathbf{x})\gamma^{ij}k_{j}\right)}{d\tau}\cdot\frac{d\tau}{d\lambda}\\
 & =\frac{d\sigma(\mathbf{x})}{d\lambda}\gamma^{ij}k_{j}\frac{d\lambda}{d\tau}\cdot\frac{d\tau}{d\lambda}+\sigma(\mathbf{x})\frac{d\gamma^{ij}}{d\lambda}k_{j}\frac{d\lambda}{d\tau}\cdot\frac{d\tau}{d\lambda}+\sigma(\mathbf{x})\gamma^{ij}\frac{dk_{j}}{d\tau}\cdot\frac{d\tau}{d\lambda}\\
 & =\frac{d\sigma(\mathbf{x})}{dx^{s}}\frac{dx^{s}}{d\lambda}\gamma^{ij}k_{j}+\sigma(\mathbf{x})\frac{\partial \gamma^{ij}}{\partial x^{s}}\frac{dx^{s}}{d\lambda}k_{j}-\sigma(\mathbf{x})\gamma^{ij}\frac{\partial H}{\partial x^{j}}\cdot\frac{d\tau}{d\lambda}\\
& =\frac{d\sigma(\mathbf{x})}{dx^{s}}k^{s}\gamma^{ij}k_{j}+\sigma(\mathbf{x})\frac{\partial\gamma^{ij}}{\partial x^{s}}k^{s}k_{j}-\sigma(\mathbf{x})\gamma^{ij}\frac{1}{2}\frac{\partial \left(\sigma(\mathbf{x})\gamma^{lm}\right)}{\partial x^{j}}k_{l}k_{m}\\
 & \quad-\sigma(\mathbf{x})\gamma^{ij}\frac{1}{2}\frac{\partial f_{A}(\mathbf{x})}{\partial
 x^{j}}\left[\sigma(\mathbf{x})\gamma^{ij}k_{i}k_{j}-\frac{\omega^{2}}{c^{2}}\right]\cdot\frac{1}{f_A(\mathbf{x})}\\
\text{combining with \ \ } &
\sigma(\mathbf{x})\gamma^{ij}k_{i}k_{j}-\frac{\omega^{2}}{c^{2}}=0\
\ \ \ \textrm{and}\ \ \ \
k_j=\frac{\gamma_{jt}}{\sigma(\mathbf{x})}k^t,\
k_l=\frac{\gamma_{ls}}{\sigma(\mathbf{x})}k^s,\
k_m=\frac{\gamma_{mt}}{\sigma(\mathbf{x})}k^t\\
 & =\gamma^{ij}\gamma_{jt}\frac{1}{\sigma(\mathbf{x})}\frac{d\sigma(\mathbf{x})}{dx^{s}}k^{s}k^{t}+\gamma_{jt}\frac{\partial \gamma^{ij}}{\partial x^{s}}k^{s}k^{t}-\frac{1}{2}\gamma_{mt}\frac{\gamma_{ls}}{\sigma(\mathbf{x})}\gamma^{ij}\frac{\partial \left(\sigma(\mathbf{x})\gamma^{lm}\right)}{\partial
 x^{j}}k^{s}k^{t}
\end{align*}
\begin{align*}
 & =\gamma^{ij}\gamma_{jt}\frac{1}{\sigma(\mathbf{x})}\frac{d\sigma(x)}{dx^{s}}k^{s}k^{t}+\left(\frac{\partial \left(\gamma_{jt}\gamma^{ij}\right)}{\partial x^{s}}-\gamma^{ij}\frac{\partial \gamma_{jt}}{\partial x^{s}}\right)k^{s}k^{t}\\
 & \quad-\frac{1}{2}\gamma^{ij}\gamma_{mt}\left(\frac{\partial \left(\frac{\gamma_{ls}}{\sigma(\mathbf{x})}\cdot\sigma(\mathbf{x})\gamma^{lm}\right)}{\partial x^{j}}-\sigma(\mathbf{x})\gamma^{lm}\frac{\partial \left(\frac{\gamma_{ls}}{\sigma(\mathbf{x})}\right)}{\partial
 x^{j}}\right)k^{s}k^{t}\\
 & =\gamma^{ij}\gamma_{jt}\frac{1}{\sigma(\mathbf{x})}\frac{d\sigma(\mathbf{x})}{dx^{s}}k^{s}k^{t}-\gamma^{ij}\frac{\partial \gamma_{jt}}{\partial x^{s}}k^{s}k^{t}+\frac{1}{2}\gamma^{ij}\gamma_{mt}\sigma(x)\gamma^{lm}\frac{\partial \left(\gamma_{ls}/\sigma(\mathbf{x})\right)}{\partial x^{j}}k^{s}k^{t}\\
 & =\gamma^{ij}\gamma_{jt}\frac{1}{\sigma(\mathbf{x})}\frac{d\sigma(x)}{dx^{s}}k^{s}k^{t}+\frac{1}{2}\sigma(\mathbf{x})\frac{d\frac{1}{\sigma(\mathbf{x})}}{dx^{j}}\gamma^{ij}\gamma_{ts}k^{s}k^{t}-\gamma^{ij}\left(\frac{1}{2}\frac{\partial \gamma_{jt}}{\partial x^{s}}+\frac{1}{2}\frac{\partial \gamma_{jt}}{\partial x^{s}}-\frac{1}{2}\frac{\partial \gamma_{st}}{\partial x^{j}}\right)k^{s}k^{t}\\
 & =-\Gamma_{st}^{i}k^{s}k^{t}+\delta_{t}^{i}\frac{1}{\sigma(\mathbf{x})}\frac{d\sigma(\mathbf{x})}{dx^{s}}k^{s}k^{t}-\frac{1}{2}\frac{1}{\sigma(\mathbf{x})}\frac{d\sigma(\mathbf{x})}{dx^{j}}\gamma^{ij}\gamma_{ts}k^{s}k^{t}\\
 &
 =-\left(\Gamma_{st}^{i}-\delta_{t}^{i}\frac{d\ln\left|\sigma(\mathbf{x})\right|}{dx^{s}}+\frac{d\ln\sqrt{\left|\sigma(\mathbf{x})\right|}}{dx^{j}}\gamma^{ij}\gamma_{ts}\right)k^{s}k^{t}.
\end{align*}
So we get the new geodesic Eq.
(\ref{eq:geodesic_equation_conformal}), where $\Gamma_{st}^{i}$ is
given by the coordinate transformation and the other terms are given
by the conformation transformation
(\ref{eq:conformal_transformation}).

\section{Four dimensional
transformation\label{sec:Four-dimensional-transformation}}\label{appendix_C}

\subsection{The proof process}\label{appendix_C_proof}

We have the Hamiltonian (\ref{eq:hamiltonian_4D_g_00}) and the
canonical equations (\ref{eq:canonical equations}). For the $k^i$'s
definition, we can derive that
\begin{align*}
k^{i} =\frac{dx^{i}}{d\lambda}
=\frac{dx^i}{d\tau}\cdot\frac{d\tau}{d\lambda} =\frac{\partial
H}{\partial k^i}\cdot\frac{1}{f_A(\mathbf{x})}
=2\frac{f'(x)}{f_A}\gamma^{il}k_{i}.
\end{align*}
Just as previous calculation, we can define
$f'(\mathbf{x})=f_{A}(\mathbf{x})/2$. Then the Hamiltonian
(\ref{eq:hamiltonian_4D_g_00}) can be written as
\begin{align}
H &
=\frac{f_{A}(\mathbf{x})}{2}\left(\gamma^{ij}k_{i}k_{j}-g^{00}k_{0}k_{0}\right).\label{eq:new_hamiltonian2}
\end{align}
Under this situation, $\gamma^{ij}$ is independent of $x^{0}$, by
using the specialized Hamiltonian(\ref{eq:new_hamiltonian2}) and the
following canonical Eq. (\ref{eq:canonical equation dkdt}), we can
derive
\begin{align*}
\frac{dk^{i}}{d\lambda} & =\frac{d\left(\gamma^{ij}k_{j}\right)}{d\lambda}\cdot\frac{d\lambda}{d\tau}\\
& =\frac{\partial \gamma^{ij}}{\partial x^m}\cdot\frac{dx^m}{d\tau}\frac{k_j}{f_A(\mathbf{x})}+\gamma^{ij}\frac{dk_j}{d\tau}\cdot\frac{1}{f_A(\mathbf{x})}\\
& =\frac{\partial \gamma^{ij}}{\partial x^m}\cdot\frac{\partial H}{\partial k_m}\frac{k_j}{f_A(\mathbf{x})}-\gamma^{ij}\frac{\partial H}{\partial x^j}\cdot\frac{1}{f_A(\mathbf{x})}\\
& =\gamma_{mn}\frac{\partial \gamma^{ij}}{\partial
x^m}k_nk_j-\frac{1}{2}\gamma^{ij}\left[\frac{\partial
\gamma^{mn}}{\partial x^j}k_mk_n-\frac{\partial g^{00}}{\partial
x^j}(k_0)^2\right]-\frac{\gamma^{ij}}{2f_A(\mathbf{x})}\frac{\partial
f_A(\mathbf{x})}{\partial
x^j}\left(\gamma^{mn}k_mk_n-g^{00}k_0k_0\right)\\
& =\gamma^{mn}\frac{\partial \gamma^{ij}}{\partial x^m}\gamma_{ns}\gamma_{jt}k^sk^t-\gamma^{ij}\frac{\partial \gamma^{mn}}{\partial x^j}\gamma^{mn}\gamma_{nt}k^sk^t+\frac{1}{2}\gamma^{ij}\frac{\partial (1/g_{00})}{\partial x^j}(k_0)^2\\
& =-\gamma^{ij}\frac{\partial \gamma_{jt}}{\partial
x^m}\delta^m_sk^tk^s+\frac{1}{2}\gamma^{ij}\frac{\partial
\gamma_{ms}}{\partial
x^j}\delta^m_tk^sk^t-\frac{1}{2g_{00}^2}\gamma^{ij}\frac{\partial
g_{00}}{\partial x^j}(g_{00})^2(k^0)^2
\end{align*}
\begin{align*}
&=-\frac{1}{2}\gamma^{ij}\left(\frac{\partial \gamma_{jt}}{\partial
x^s}+\frac{\partial \gamma_{js}}{\partial x^t}-\frac{\partial
\gamma_{st}}{\partial
x^j}\right)k^sk^t-\frac{1}{2}\gamma^{ij}\frac{\partial
g_{00}}{\partial x^j}(k^0)^2\\
\text{because } & \Gamma^i_{jk}=\frac{1}{2}g^{i\mu}\left(g_{j\mu,k}+g_{k\mu,j}-g_{jk,\mu}\right)=\frac{1}{2}\gamma^{ij}\left(\frac{\partial \gamma_{jt}}{\partial x^s}+\frac{\partial \gamma_{js}}{\partial x^t}-\frac{\partial \gamma_{st}}{\partial x^j}\right)\\
\text{and } & \Gamma_{00}^{i}=\frac{1}{2}g^{ij}\left(g_{j0,0}+g_{j0,0}-g_{00,j}\right)=-\frac{1}{2}g^{ij}g_{00,j}=\frac{1}{2}\gamma^{ij}g_{00,j}\\
\text{and } & \Gamma_{t0}^{i}=\frac{1}{2}g^{ij}\left(g_{jo,t}+g_{jt,o}-g_{to,j}\right)=0\\
\text{So we finally get}&\\
\frac{dk^i}{d\lambda}&=-\Gamma_{st}^{i}k^{s}k^{t}-\Gamma_{00}^{i}k^{0}k^{0}-\Gamma_{t0}^{i}k^{0}k^{t}-\Gamma_{0t}^{i}k^{t}k^{0}=-\Gamma^i_{\mu\nu}k^{\mu}k^{\nu}
\end{align*}
which is the geodesic equation in the 4-dimensional space-time.

\subsection{Comparing with conformal
transformation}\label{appendix_C_compare}

Furthermore, because\[ds^{2}=g_{ts}k^{s}k^{t}+g_{00}k^{0}k^{0}=0,\]
we have
\[g_{00}k^{0}k^{0}=-g_{ts}k^{s}k^{t}=\gamma_{ts}k^sk^t.\]
Thus
\begin{eqnarray}\label{eq:52}
\frac{dk^{i}}{d\lambda}+\left(\Gamma_{st}^{i}k^{s}k^{t}+\frac{1}{2}\gamma^{ij}g_{00,j}k^{0}k^{0}\right)&=0\nonumber \\
\frac{dk^{i}}{d\lambda}+\left(\Gamma_{st}^{i}k^{s}k^{t}+\frac{1}{2}\gamma^{ij}\frac{\partial g_{00}}{\partial x^{j}}\frac{1}{g_{00}}\gamma_{st}k^{s}k^{t}\right)&=0\nonumber \\
\frac{dk^{i}}{d\lambda}+\left(\Gamma_{st}^{i}+\frac{\partial\ln\sqrt{g_{00}}}{\partial
x^{j}}\gamma^{ij}\gamma_{st}\right)k^{s}k^{t}&=0.\label{eq:canonical2geodesic_4d}
\end{eqnarray}
If we make $\sigma(\mathbf{x})=g_{00}$ in the equation
(\ref{eq:geodesic_equation_conformal})
\begin{equation}\label{eq:conformal_geodesic_goo}
\frac{dk^i}{d\lambda}+\left(\Gamma^i_{st}+\frac{\partial\ln\sqrt{g_{00}}}{\partial
x^j}\gamma^{ij}\gamma_{st}-\delta_t^i\frac{\partial\ln|g_{00}|}{\partial
x^s}\right)k^sk^t=0,
\end{equation}
and compare it with the equation (\ref{eq:52}), there is one more
item $-\delta_t^i\frac{\partial\ln|g_{00}|}{\partial x^s}$ in the
equation (\ref{eq:conformal_geodesic_goo}), which shows different
geodesic lines for two geodesic equations deriving from two kinds of
transformations.

\end{appendix}

\end{document}